\documentclass[titlepage,12pt,a4paper,reqno]{article}
\usepackage{amsmath,amssymb,amsfonts,graphics}
\usepackage[linktocpage=true,colorlinks,pdftex]{hyperref}
\usepackage{xcolor}
\colorlet{linkequation}{blue}
\usepackage{bm}
\usepackage{doi}
\usepackage{hyperref}
\hypersetup{colorlinks, citecolor=violet, filecolor=black, linkcolor=black, urlcolor=blue}
\newcommand*{\refeq}[1]{%
  \begingroup
    \hypersetup{
      linkcolor=linkequation,
      linkbordercolor=linkequation,
    }%
    \ref{#1}%
  \endgroup
}
\allowdisplaybreaks

\begin{document} 


\begin{titlepage}

\centerline{\Large \bf Backreaction of particle production on physical gravitons} 
\medskip
\centerline{\Large \bf in the de Sitter space}
\vskip 1cm
\centerline{ \bf Ion V. Vancea }
\vskip 0.5cm
\centerline{\sl Grupo de F{\'{\i}}sica Te\'{o}rica e F\'{\i}sica Matem\'{a}tica}
\centerline{\sl Departamento de F\'{\i}sica}
\centerline{\sl Universidade Federal Rural do Rio de Janeiro}
\centerline{\sl Cx. Postal 23851, BR 465 Km 7, 23890-000 Serop\'{e}dica - RJ,
Brasil}
\centerline{
\texttt{\small ionvancea@ufrrj.br} 
}

\vspace{0.5cm}

\centerline{20 March 2019}

\vskip 1.4cm
\centerline{\large\bf Abstract}

We derive the effects of the scalar particle production on the physical graviton operator in the de Sitter space. Our analysis is done for the sub-horizon modes at large values of the conformal time. In this limit, we completely determine the correction to the scalar energy-momentum tensor at the first order in the WKB iteration. Also, we calculate the corresponding correction to the graviton operator for a single sub-horizon mode.

\vskip 0.7cm 

 Quantum field theory in the de Sitter space. Quantum gravity in the de Sitter space. 

\noindent

\end{titlepage}


\section{Introduction}

The results of the astronomical observations of the Type 1a supernovae that are consistent with an accelerated expansion of the universe \cite{Riess:1998cb,Perlmutter:1998np} have been a strong motivation for the renewed interest in the properties of matter and gravity in the de Sitter space. Other catalysts for the development of the quantum field theories on the de Sitter space are the generalization of the duality between the string gravity on the anti-de Sitter space and the conformal quantum field theories \cite{Maldacena:1997re} and the research on the microscopic structure of the Dark Matter. The quantum matter in curved space-times has distinct features from the Minkowski space-time that are due mainly to the interaction with the gravitational background. One such new process is the particle production which, when analysed in the de Sitter space, could give interesting information about the quantum phenomena that took place in the primordial universe. This important subject has been studied for a long time in the literature starting with \cite{Parker:1969au,Parker:1971pt} and can be found in several standard texts on quantum field theory on curved manifolds, see e. g.  
\cite{Birrell:1982ix,Mukhanov:1990me,Weinberg:2008zzc} and \cite{Parker:2012at,Parker:2017imt} for an historical overview. 

The creation of particles in the gravitational background can be viewed as a quantum backreaction process since it changes the matter energy-momentum tensor which is the source of the gravitational field. In the limit of the weak fields, the particle production can be studied with the canonical methods and differential equations can be established for the quantum operators. 

In the present letter, we are going to study the effect of the scalar particle creation on the physical graviton operator in the de Sitter space. Our main focus are the sub-horizon modes of the scalar field in the early universe for which we determine the correction to the energy-momentum tensor at the first order in the WKB expansion between two instantaneous vacua defined in terms of the conformal time. Although the de Sitter space is restrictive enough to fix a unique vacuum for all the modes of the scalar field, namely the Bunch-Davies vacuum, it is interesting to have the system prepared in instantaneous vacua, 
for example in the scalar field inflation where there are small changes in energy at minima of the slow-rolling potential \cite{Louis:2018btq} or in the investigation of the contribution of the neutrino mixing to the cosmological constant \cite{Blasone:2004td}. By determining the effect of the particle creation on the energy-momentum tensor from an anterior instant, we are  
able to derive the correction to the physical graviton operator which represents our main result. The analysis presented here can be performed for other vacua, including the Bunch-Davies and the asymptotic vacua, respectively. In fact, we give the first order correction to the energy-momentum tensor in its general form that can be applied immediately to these cases. The results showed here are interesting for potential applications to the phenomena mentioned above as well as to the study of the interaction between the Dark Matter particles and the gravitons in the primordial universe within the simplified Dark Matter models with graviton mediators approach \cite{Vancea:2018bom,Vancea:2018hqp}. The latter represents our main interest in this problem. 

The same problem has been addressed in the literature in the Schwinger-Keldysh (SK) formalism in \cite{Park:2011ww,Park:2011kg,Park:2015kua} and previously in \cite{Allen:1987tz} for the massless, minimally coupled scalar. In the last paper, the vacuum expectation value of the stress tensor was shown to be exactly a constant that can be absorbed into the renormalized cosmological constant. The calculations in the SK formalism from \cite{Park:2011ww,Park:2011kg} also predict no modification of the graviton self-energy at one loop from the massless scalar fields. The corrections are present only for the gravitational potentials as shown in \cite{Park:2015kua}. Since one expects that the massless fields generate the largest corrections to the stress tensor, 
the results given by the WKB obtained here, and by the SK formalism from the literature, respectively, produce a apparent tension in the massless limit whose resolution is beyond the scope of the present work whose main objective is to determine the correction to the graviton operators in the WKB approach.

This paper is organized as follows. In Section 2, we review some results on the graviton and the scalar field in the physical gravitational gauge in the de Sitter space as given in \cite{Ford:1977dj}-\cite{Faizal:2011iv}. Also, we establish our notations that are close to the ones used in \cite{Vancea:2018hqp}. In Section 3, we calculate the correction to the energy-momentum tensor of the scalar field from the particle production of a single sub-horizon mode. From that, we derive the effect of this correction on the graviton operator. The last section is devoted to discussions. There, we briefly comment on the results obtained here in the WKB formalism and the one calculated in the SK formalism. Throughout this work we use the natural units with $c= 16 \pi G = 1$.

\section{Physical gravitons in the de Sitter space}

In this section, we briefly review some properties of the physical gravitons in the de Sitter space and establish our notations. Most of the information on gravitons can be found in the standard texts on quantum field theory in curved space-time, e. g. \cite{Birrell:1982ix,Mukhanov:1990me} and on the physical gravitational gauge in the papers \cite{Ford:1977dj} - \cite{Faizal:2011iv}.  

The physical processes that took place in the early universe during inflation and the preheating phase produced quantum excitations of the space-time metric which in the linear approximation define the gravitons \cite{Mukhanov:1990me}. The classical graviton field $h_{\mu \nu}$ with $|h_{\mu \nu}| << 1$ is defined as a linear fluctuation of the background metric $g_{\mu \nu}$ which is an exact solution to the Einstein's equations. The dynamics of the free graviton field is given by the following action
\begin{align}
S_0[h] & =  \int d^4 x \, 
\sqrt{-g} 
\left[ 
\frac{1}{2} 
\nabla_{\mu} h^{\mu \rho} \nabla^{\nu} h_{\nu \rho}
- \frac{1}{4} \nabla_{\mu} h_{\nu \rho} \nabla^{\mu} h^{\nu \rho} 
\right. 
\nonumber \\
&
\left.
+ \frac{1}{4} 
\left( 
\nabla^{\mu} h - 2 \nabla^{\nu} {h^{\mu}}_{\nu} 
\right)
\nabla_{\mu} h
- \frac{1}{2}
\left(
h_{\mu \nu} h^{\mu \nu} 
+ \frac{1}{2} h^2 
\right)
\right] 
\, , 
\label{graviton-free-action}
\end{align}
where $h = {h^{\mu}}_{\mu}$. Not all the components of the tensor $h_{\mu \nu}$ represent physical degrees of freedom since the action (\refeq{graviton-free-action}) is invariant under the gravitational gauge transformations
\begin{equation}
h_{\mu \nu} \rightarrow {h'}_{\mu \nu} = h_{\mu \nu} + 
\nabla_{\mu} \xi_{\nu} + \nabla_{\nu} \xi_{\nu} 
\, ,
\label{gravitational-gauge-transformations}
\end{equation}
where $\xi_{\mu}$ is an arbitrary smooth vector field. The non-physical degrees of freedom can be eliminated by choosing the physical gauge in which the graviton is transverse and traceless. These conditions are imposed by the following equations \cite{Ford:1977dj}
\begin{equation}
\nabla^{\mu} h_{\mu \nu} = 0
\, ,
\qquad
{h^{\mu}}_{\mu} = 0
\, ,
\qquad
n^{\mu} h_{\mu \nu} = 0
\, ,
\label{transverse-traceless}
\end{equation}
where the last relation defines the synchronous graviton and $n^{\mu}$ is the unit vector parallel to $\partial/\partial \tau$. 

One maximally symmetric solution to the Einstein's equation that describes an accelerated expansion phase of the primordial universe is the de Sitter space. In the conformal time coordinates
\begin{equation}
ds^2 = g_{\mu \nu} (t) \, d x^{\mu} d x^{\nu} = - \frac{1}{H_0 \tau} \eta_{\mu \nu} dx^{\mu} dx^{\nu} \, , 
\label{de-Sitter-conformal-metric}
\end{equation}
where $\tau \in (-\infty , 0 ]$ is the conformal time parameter and $H_0$ is the Hubble's constant\footnote{Since the evolution should be described with an increasing time parameter, we will consider whenever necessary $|\tau|$ instead of $\tau$ without changing the notation. Alternatively, one can cover the future light-cone of the Penrose diagram with the parameter $\tau \in [0, \infty )$.}. Then the equation of motion of the free physical graviton derived from the action (\refeq{graviton-free-action}) in the de Sitter background (\refeq{de-Sitter-conformal-metric}) takes the following simple form
\begin{equation}
\left( 
\Box - 2 H_{0}^{2} 
\right) h_{\mu \nu}
= 0
\, .
\label{eq-motion-phyiscal-graviton-de-Sitter}
\end{equation}
The equation (\refeq{eq-motion-phyiscal-graviton-de-Sitter}) has been exhaustively studied in the literature in many places. Since the non-physical degrees of freedom have been removed, the classical graviton field can be quantized by using the canonical methods (see e. g. \cite{Ford:1977dj}).

The free graviton field describes the excitations of the de Sitter metric away from the physical sources. These can be included by applying the linearization procedure to the Einstein's equations with matter which is represented by the energy-momentum tensor $T^{(mat)}_{\mu \nu}$. If one intends to investigate quantum phenomena, both the graviton field and the matter component should be quantized in the classical de Sitter background.

Consider the minimally coupled scalar field $\phi$ in the de Sitter space  characterized by the general action
\begin{equation}
S_0 [g, \phi] =  \int d^4 x \sqrt{-g} \,
\left[
\frac{1}{2} g^{\mu \nu} \, \nabla_{\mu} \phi \, \nabla_{\nu} \phi 
- V(\phi) 
\right] \, ,
\label{action-scalar}
\end{equation}
where $g$ denotes the background metric from the equation (\refeq{de-Sitter-conformal-metric}) and $V(\phi )$ is an arbitrary potential smooth function on $\phi$. The energy-momentum tensor calculated from $S_0 [g, \phi]$ has the following form
\begin{equation}
T^{(\phi)}_{\mu \nu} = \partial_{\mu} \phi \partial_{\nu} \phi - 
g_{\mu \nu} 
\left[
\partial^{\rho} \phi \partial_{\rho} \phi - V(\phi)
\right] \, .
\label{scalar-energy-momentum}
\end{equation}
If we consider the field $\phi$ as the source of the physical graviton excitations of the de Sitter metric, the homogeneous equation (\refeq{eq-motion-phyiscal-graviton-de-Sitter}) becomes
\begin{equation}
\left( 
\Box - 2 H_{0}^{2} 
\right) h_{\mu \nu}
= T^{(\phi)}_{\mu \nu}
\, .
\label{eq-motion-physical-graviton-de-Sitter-1}
\end{equation}
Some important remarks are in order here. According to the gauge fixing conditions from the equation (\refeq{transverse-traceless}), the metric is excited along the spatial directions only. Moreover, the synchronous and traceless constraints together imply that $h_{0i} = h_{00} = h_{33} = 0$. Therefore, the only components of $ T^{(\phi)}_{\mu \nu}$ that source these excitations are $ T^{(\phi)}_{ij}$. Thus, the equation (\refeq{eq-motion-physical-graviton-de-Sitter}) is sometimes written as
\begin{equation}
\left( 
\Box - 2 H_{0}^{2} 
\right) h^{TT}_{i j}
= T^{(\phi)TT}_{i j}
\, ,
\label{eq-motion-physical-graviton-de-Sitter}
\end{equation}
where $T^{(\phi)TT}_{i j}$ denotes the transverse-traceless spatial components of the energy-momentum tensor. To simplify the notations, we will drop the $TT$ superscripts in what follows but keep in mind that the tensors are of transverse-traceless type.

The quantization of the theory can be performed as usual by applying the canonical quantization method (see e. g. \cite{Birrell:1982ix}). In order to quantize the field $\phi$, some assumptions should be made on its parameters and on the potential $V(\phi )$ from the relation (\refeq{action-scalar}). In what follows, we are going to consider that $V(\phi ) = m^{2}_{eff} (\tau ) \phi^2$ corresponds to a free field of effective de Sitter mass
\begin{equation}
m^{2}_{eff} (\tau ) = 
\left(
\frac{m^2}{H^{2}_{0}}
- 2
\right) \frac{1}{\tau^2} 
\, .
\label{scalar-effective-mass-dS}
\end{equation}
Since most of the particles had masses much smaller than the Hubble's constant in the early universe, we will take further $m^{2}_{eff} (\tau ) \simeq - 2 \tau^{-2}$. 

In the conformal coordinates, the de Sitter patch is conformally mapped into the Minkowski space. Therefore, the free graviton and the scalar field admit Fourier decompositions in the spatial directions of the following form
\begin{align}
h_{ij} (\tau, \mathbf{x} )  & = 
\int \frac{d^3 \mathbf{k}}{(2 \pi)^3}
\mathrm{F}_{ij}(\tau, \mathbf{k}) \, 
e^{i \mathbf{k} \mathbf{x}} 
\, ,
\label{Fourier-expansion-graviton}
\\
\phi(\tau, \mathbf{x} ) & = \int \frac{d^3 \mathbf{k}}{(2 \pi)^3} \, 
\varphi(\tau , \mathbf{k}) \, e^{i \mathbf{k} \mathbf{x}} 
\, .
\label{Fourier-expansion-scalar}
\end{align}
The mode functions $\mathrm{F}_{ij}(\tau, \mathbf{k})$ and $\varphi(\tau , \mathbf{k})$ satisfy the following equations of motion \cite{Vancea:2018hqp}
\begin{align}
\left[
\partial^{2}_{\tau} 
-
\frac{2}{\tau } \partial_{\tau}
+
\mathbf{k}^2
\right]
\mathrm{F}_{ij}(\tau, \mathbf{k}) 
& = 
\frac{1}{ H^{2}_{0} \tau^2} 
\mathrm{T}^{(\phi)}_{ij}(\tau, \mathbf{k}) \, ,
\label{equation-motion-grav-mode}
\\
\partial^{2}_{\tau} \mathrm{\phi}(\tau , \mathbf{k} ) + \omega^{2}_{\mathbf{k}} (\tau )
 \mathrm{\phi} (\tau , \mathbf{k} ) & = 0 \, ,
\label{scalar-SM-scalar-mode}
\end{align}
where we recognize the equation (\refeq{scalar-SM-scalar-mode}) as the Sasaki-Mukhanov equation \cite{Sasaki:1986hm,Mukhanov:1988jd} and $\omega^{2}_{\mathbf{k}} (\tau ) = k^2 + m^{2}_{eff} (\tau )$. The right hand side of the equation (\refeq{equation-motion-grav-mode}) contains the Fourier modes of the scalar field energy-momentum tensor defined by the following expansion \cite{Vancea:2018hqp}
\begin{equation}
T^{(\phi)}_{ij} (\tau, \mathbf{x} )   = 
\int \frac{d^3 \mathbf{k}}{(2 \pi)^3} \,
\mathrm{T}^{(\phi)}_{ij}(\tau, \mathbf{k}) \, 
e^{i \mathbf{k} \mathbf{x}} 
\, .
\label{energy-momentum-modes}
\end{equation}
The homogeneous equation (\refeq{equation-motion-grav-mode}) and the equation (\refeq{scalar-SM-scalar-mode}) can be solved exactly and their general solution can be put into the canonical form as follows
\begin{align}
h_{0,ij} (\tau, \mathbf{x} )  & = 
\frac{1}{2}
\int \frac{d^3 \mathbf{k}}{(2 \pi)^3}
\sum_{\sigma = 1, 2}  \, 
\tau^{\frac{3}{2}} \,
\left\{
a^{(\sigma)} (\mathbf{k}) H^{(1)}_{\frac{3}{2}} (k \tau ) 
e^{i \mathbf{k} \mathbf{x}} 
+
a^{(\sigma) \dagger} (\mathbf{k}) H^{(2)}_{\frac{3}{2}} (k \tau ) 
e^{-i \mathbf{k} \mathbf{x}}
\right\}
\epsilon^{(\sigma)}_{ij} (\mathbf{k})
\, ,
\label{physical-graviton-Fourier-exp}
\\
\mathrm{\phi} (\tau , \mathbf{k}) & = 
\frac{1}{\sqrt{2}}
\left[
b_{\mathbf{k}} \mathcal{\phi}^{*}_{\mathbf{k}} (\tau ) +
b^{+}_{-\mathbf{k}} \mathcal{\phi}_{\mathbf{k}} (\tau )
\right] \, .
\label{scalar-mode-phi}
\end{align}
Here, $ H^{(1,2)}_{\frac{3}{2}} (k \tau )$ denote the Hankel's functions and $\epsilon^{\sigma}_{ij} (\mathbf{k}) $,  $\sigma =  1, 2 = + , \times $ is the polarization tensor. Upon quantization, the Fourier coefficients become mode operators that satisfy the canonical commutation relations
\begin{align}
\left[
a^{\sigma}(\mathbf{p}) ,
a^{\lambda \dagger}(\mathbf{k})
\right] & = (2 \pi)^3 \delta^{\sigma \lambda} \, \delta ( \mathbf{p} - \mathbf{k}) 
\, ,
\label{graviton-commutators}
\\
\left[
b_{\mathbf{p}} , b^{+}_{\mathbf{k}}
\right] & = \delta (\mathbf{p} - \mathbf{k}) \, ,
\label{DM-scalar-commutator}
\end{align}
the rest of the commutators being zero as usual. The time-independent vacuum of the system $| \Omega \rangle$ is annihilated by all the annihilation operators
\begin{equation}
a^{\sigma}(\mathbf{k}) | \Omega \rangle = 
b (\mathbf{k}) | \Omega \rangle =
0 \, ,
\qquad
\forall \sigma = 1, 2 \, , \, \, \, \forall \mathbf{k} \, \in \mathbb{R}^3 
\, .
\label{graviton-vacuum}
\end{equation}
Note that the scalar mode functions $\varphi_{\mathbf{k}} (\tau )$ are linearly independent solutions of the Sasaki-Mukhanov equation (\refeq{scalar-mode-phi}). However, in order to be determined completely, further conditions that reflect the particular physical phenomena under study should be imposed on $\varphi_{\mathbf{k}} (\tau )$'s. These conditions can be used to select a particular vacuum of the theory. Nevertheless, it might occur that not all modes of the scalar field be able to satisfy a given set of conditions. In particular, the de Sitter space is restrictive enough to have a unique vacuum for all the modes, the Bunch-Davies vacuum. However, it is interesting to have the system prepared in different vacua in order to study the phenomena that relate locally closed events.

\section{Effect of scalar particle production on gravitons}

In the quantum theory, the set of equations (\refeq{equation-motion-grav-mode}) 
and (\refeq{scalar-SM-scalar-mode}) 
allow one to calculate the physical graviton field in terms of its quantum source represented by the quantized scalar field at any time. However, in the gravitational de Sitter background, there is a backreaction due to the scalar particle production (see e. g. \cite{Weinberg:2008zzc} ). In order to estimate the effect of the scalar particle creation on the graviton, we consider the instantaneous vacua at two close instants $0 << \tau_1 < \tau_2$ in the early universe. The existence of the instantaneous vacua requires that $\omega^{2}_{\mathbf{k}} (\tau ) > 0$ which is true for the sub-horizon modes for which $k \tau >> 1$\footnote{Here $\tau$ stands for the modulus of the conformal time}. 

Let us determine the effect of particles creation for a single sub-horizon mode $\mathbf{k}$. Since $\tau_1$ and $\tau_2$ are close to each other, the physical conditions do not change significantly and the same description applies to the field $\phi$ at both times. It follows that there is a linear Bogoliubov map between the Hilbert spaces $\mathcal{H}_{\mathbf{k}}(\tau_1 )$ and $\mathcal{H}_{\mathbf{k}}(\tau_2)$ that induces a map on the corresponding spaces of operators $\mathsf{End}[\mathcal{H}_{\mathbf{k}}(\tau_1 )]$ and $\mathsf{End}[\mathcal{H}_{\mathbf{k}}(\tau_2 )]$.
This map is characterized by the time-dependent Bogoliubov coefficients
$\alpha_{\mathbf{k}}(\tau_1 , \tau_2 )$ and $\beta_{\mathbf{k}}(\tau_1 , \tau_2 )$. The normalization condition of the mode functions $\varphi_{\mathbf{k}} (\tau_1 )$ and $\tilde{\varphi}_{\mathbf{k}} (\tau_2 )$, respectively, related by the Bogoliubov transformation, requires that the Bogoliubov coefficients satisfy the following equation
\begin{equation}
|\alpha_{\mathbf{k}}(\tau_1 , \tau_2 )|^2 - |\beta_{\mathbf{k}}(\tau_1 , \tau_2 )|^2 = 1 \, .
\label{Bogoliubov-coeff-normalization}
\end{equation}
Consider the sub-horizon mode functions of the instantaneous vacuum at $\tau_1$ of the following form
\begin{equation}
\varphi_{\mathbf{k}} (\tau_1 ) = \frac{1}{\sqrt{k}} e^{i k \tau_1} \, ,
\label{DM-instantaneous-mode-fct-1}
\end{equation}
which is the same as in the Minkowski space-time. At the instant $\tau_2$, the corresponding mode function can be obtained by applying the Bogoliubov map and can be written as follows
\begin{equation}
\tilde{\varphi}_{\mathbf{k}} (\tau_2 ) = \alpha_{\mathbf{k}}(\tau_1 , \tau_2 ) \varphi_{\mathbf{k}} (\tau_1 ) +
\beta_{\mathbf{k}}(\tau_1 , \tau_2 ) \varphi^{*}_{\mathbf{k}} (\tau_1 ) \, .
\label{DM-instantaneous-mode-fct-2}
\end{equation}
In order to determine $\tilde{\varphi}_{\mathbf{k}} (\tau_2 )$, one has to find firstly the Bogoliubov coefficients.  In the WKB approximation, they can be calculated in terms of the auxiliary function $\zeta_{\mathbf{k}} (\tau )$ that is a solution of the following problem
\begin{equation}
\frac{d \zeta_{\mathbf{k}} (\tau )}{d \tau} = 
- 2 \, i \, \omega_{\mathbf{k}}(\tau ) \zeta_{\mathbf{k}} (\tau )
+ \frac{1}{2 \omega_{\mathbf{k}}(\tau )}
\left[ 
1 - \zeta_{\mathbf{k}}^2 (\tau )
\right]
\frac{d \omega_{\mathbf{k}} (\tau )}{d \tau}  \, ,
\qquad
\zeta_{\mathbf{k}} (\tau_1 ) = 0 \, .
\label{zeta-function}
\end{equation}
Then the Bogoliubov coefficients are given by the following relations
\begin{align}
\alpha_{\mathbf{k}}(\tau_1 , \tau_2 ) & = 
\left[
 1 - |\zeta_{\mathbf{k}} (\tau_1 , \tau_2 ) |^2
\right]^{-\frac{1}{2}} \, ,
\label{alpha-zeta}
\\
\beta_{\mathbf{k}}(\tau_1 , \tau_2)  & = 
\zeta^{*}_{\mathbf{k}} (\tau_1 , \tau_2 )
\left[
 1 - |\zeta_{\mathbf{k}} (\tau_1 , \tau_2 ) |^2
\right]^{-\frac{1}{2}} \, .
\label{beta-zeta}
\end{align}
The equation (\refeq{zeta-function}) can be solved iteratively. In what follows, we are going to limit ourselves to the first order in the iteration at which the problem given by the equation  (\refeq{zeta-function}) takes the following form
\begin{equation}
\frac{d \zeta^{(1)}_{\mathbf{k}} (\tau )}{d \tau} = 
- 2 \, i \, \omega_{\mathbf{k}}(\tau ) \zeta^{(1)}_{\mathbf{k}} (\tau )
+ \frac{1}{2 \omega_{\mathbf{k}}(\tau )}
\frac{d \omega_{\mathbf{k}} (\tau )}{d \tau}  \, ,
\qquad
\zeta^{(1)}_{\mathbf{k}} (\tau_1 ) = 0 \, .
\label{zeta-function-1}
\end{equation}
From this, one deduces that the general solution to the equation 
(\refeq{zeta-function-1}) is given by the relation
\begin{equation}
\zeta^{(1)}_{\mathbf{k}} (\tau_2 , \tau_1 ) =
\int^{\tau_2}_{\tau_1} d \tau \frac{1}{2 \omega_{\mathbf{k}}(\tau )}
\frac{d \omega_{\mathbf{k}} (\tau )}{d \tau} 
\exp
\left[
- 2i \int^{\tau_2}_{\tau} d \tau' \, \omega_{\mathbf{k}}(\tau' ) 
\right]
\, .
\label{zeta-integral}
\end{equation}
In order to resolve the right hand side of the equation (\refeq{zeta-integral}) we calculate the inner integral first. 
By applying elementary methods, we obtain its exact solution of the form
\begin{equation}
\left.
\left[
\sqrt{k^2 \tau^{'2} - 2 } 
+ \sqrt{2} \cot^{-1} 
\sqrt{
\left( 
\frac{k^2 \tau^{'2}}{2} - 1
\right)}
\right]
\right|^{\tau_2}_{\tau} 
\, .
\label{solution-integral-exp}
\end{equation}
For the sub-horizon modes, the constants become irrelevant and the $\cot^{-1}(x)$ can be approximated by its asymptotic value. Then the equation (\refeq{zeta-integral}) is reduced to the following expression
\begin{equation}
\zeta^{(1)}_{\mathbf{k}} (\tau_2 , \tau_1 ) \simeq
\frac{1}{2} e^{-2ik\tau_2}
\int^{\tau_2}_{\tau_1} d \tau \, \frac{e^{2ik\tau}}{\omega_{\mathbf{k}}(\tau )}
\frac{d \omega_{\mathbf{k}} (\tau )}{d \tau} 
\, .
\label{zeta-integral-1}
\end{equation}
Here, the symbol $\simeq$ denotes that the right hand side of the equation (\refeq{zeta-integral-1}) is given in the approximations mentioned above.  
Some more calculations show that the integral from the equation (\refeq{zeta-integral-1}) can be solved in terms of the exponential integral $\mathrm{E}_{1}(z)$ with the following result
\begin{equation}
\frac{1}{4}
\left.
\left\{
e^{2i\sqrt{2}} 
\mathrm{E}_{1}[-2i(\sqrt{2} - k\tau)]
-
2 \mathrm{E}_{1}(2ik\tau)
+
e^{-2i\sqrt{2}}
\mathrm{E}_{1}[2i(\sqrt{2} + k\tau)]
\right\}
\right|^{\tau_2}_{\tau_1}
\, .
\label{solution-integral-last}
\end{equation}
The expression from the equation (\refeq{solution-integral-last}) represents a good approximation of the auxiliary function in the sub-horizon limit. After some algebra and by using the basic properties of the exponential integral \cite{Lebedev:1965}, we obtain the final form of the auxiliary function at first order
\begin{equation}
\zeta^{(1)}_{\mathbf{k}} (\tau_2 , \tau_1 ) \simeq
- \left[
\mathrm{E}_{1}(2i\tau_2) - \mathrm{E}_{1}(2i\tau_1)
\right]
e^{-2ik\tau_2}
\, .
\label{zeta-final-1}
\end{equation}
From that, one can easily write the Bogoliubov coefficients as
\begin{align}
\alpha^{(1)}_{\mathbf{k}}(\tau_1 , \tau_2 ) & \simeq
\frac{1}{\sqrt{
1 - 
\left|
\mathrm{E}_{1}(2i\tau_2) - \mathrm{E}_{1}(2i\tau_1)
\right|^2
}}
\, ,
\label{alpha-zeta-1}
\\
\beta^{(1)}_{\mathbf{k}}(\tau_1 , \tau_2)  & \simeq
\frac{
- \left[
\mathrm{E}_{1}(2i\tau_2) - \mathrm{E}_{1}(2i\tau_1)
\right]^{*}
e^{2ik\tau_2}}
{\sqrt{
1 - 
\left|
\mathrm{E}_{1}(2i\tau_2) - \mathrm{E}_{1}(2i\tau_1)
\right|^2
}}
\, .
\label{beta-zeta-1}
\end{align}
Note that the above coefficients satisfy the normalization condition given by the equation (\refeq{Bogoliubov-coeff-normalization}). Together with the mode function from the equation (\refeq{DM-instantaneous-mode-fct-1}), the Bogoliubov coefficients from the equation (\refeq{beta-zeta-1}) uniquely define the mode functions at $\tau_2$ according to the equation (\refeq{DM-instantaneous-mode-fct-2}). If we write the scalar mode fields at $\tau_2$ as
\begin{equation}
\varphi (\tau_2 , \mathbf{k}) = 
\frac{1}{\sqrt{2}}
\left[
c_{\mathbf{k}} \tilde{\varphi}^{*}_{\mathbf{k}} (\tau_2 ) +
c^{+}_{-\mathbf{k}} \tilde{\varphi}_{\mathbf{k}} (\tau_2 )
\right] \, ,
\label{DM-scalar-mode-X-2}
\end{equation}
we can see that the Bogoliubov map acts on the mode operators as 
\begin{align}
c_{\mathbf{k}} & =
\frac{1}{\sqrt{
1 - 
\left|
\mathrm{E}_{1}(2i\tau_2) - \mathrm{E}_{1}(2i\tau_1)
\right|^2
}}
\left\{
b_{\mathbf{k}} 
+
b^{+}_{\mathbf{k}}
\left[
\mathrm{E}_{1}(2i\tau_2) - \mathrm{E}_{1}(2i\tau_1)
\right]^{*}
e^{2ik\tau_2}
\right\}
\, ,
\label{c-op-1}
\\
c^{+}_{\mathbf{k}} & =
\frac{1}{\sqrt{
1 - 
\left|
\mathrm{E}_{1}(2i\tau_2) - \mathrm{E}_{1}(2i\tau_1)
\right|^2
}}
\left\{
\left[
b_{\mathbf{k}}
\mathrm{E}_{1}(2i\tau_2) - \mathrm{E}_{1}(2i\tau_1)
\right]
e^{-2ik\tau_2}
\,
+
b^{+}_{\mathbf{k}}
\right\}
\, .
\label{c-op-2}
\end{align}
The operators $c_{\mathbf{k}}$ and $c^{+}_{\mathbf{k}}$ satisfy the canonical commutation relations (\refeq{DM-scalar-commutator}) and they are interpreted as the annihilation and creation operators of the $\mathbf{k}$-mode at time $\tau_2$ and they can be used to specify the scalar field operator.

In order to evaluate the correction obtained from the scalar sub-horizon mode
$\mathbf{k}$ to the graviton field operator we ignore the effects of the graviton production between the $\tau_1$ and $\tau_2$. Recall that the graviton is in the transverse-traceless gauge. 
As such, the pure trace term from $T^{(\phi)}_{i j}$ is cancelled out and the only remaining contribution to the equation (\refeq{equation-motion-grav-mode}) is from the product $\partial_i \phi \partial_j \phi$. In terms of modes, it takes the following form
\begin{equation}
\mathrm{T}^{(\phi)}_{ij} (\tau , \mathbf{k} ) = -  
\int d^3 p \, p_i \left( k_j - p_j \right)
\varphi (\tau , \mathbf{p}) 
\varphi (\tau , \mathbf{k} - \mathbf{p}) \, . 
\label{scalar-energy-momentum-final}
\end{equation}
Therefore, the central piece to be calculated is the vacuum expectation value of $\mathrm{T}^{(\phi)}_{ij} (\tau_2 , \mathbf{k} )$ given by the equation (\refeq{scalar-energy-momentum-final}) in the vacuum at the instant $\tau_1$. 

The computations are somewhat lengthy but quite straightforward and the result can be given in a general form that is independent on the specific vacuum chosen to study the particle production. To establish it, only the general properties of the Bogoliubov transformations of the mode operators and the mode functions are needed. Since the result is quite general, we are presenting it here  
\begin{align}
\mathrm{T}^{(\phi)}_{ij} (\tau_2 , \mathbf{k} )& = 
(2 \pi )^{3} k_i k_j 
\left[
\alpha_{\mathbf{k}}(\tau_1 , \tau_2 )
\alpha^{*}_{-\mathbf{k}}(\tau_1 , \tau_2 )
+
\beta_{\mathbf{k}}(\tau_1 , \tau_2) 
\beta^{*}_{-\mathbf{k}}(\tau_1 , \tau_2) 
\right]
\nonumber
\\
& \times
\left[
\tilde{\varphi}^{*}_{\mathbf{k}} (\tau_2 )
\tilde{\varphi}_{-\mathbf{k}} (\tau_2 )
+
\tilde{\varphi}_{\mathbf{k}} (\tau_2 )
\tilde{\varphi}^{*}_{-\mathbf{k}} (\tau_2 )
\right]
\, .
\label{DM-energy-momentum-t2-general}
\end{align}
The relation (\refeq{DM-energy-momentum-t2-general}) can be particularized to the sub-horizon mode $\mathbf{k}$ whose mode function is symmetric under the reflection of the momentum and whose Bogoliubov coefficients are given by the equations (\refeq{alpha-zeta-1}) and (\refeq{beta-zeta-1}). Also, we use the fact that $\tau_1$ and $\tau_2$ are very large as to fit the early universe description. Then some algebra shows that the difference of the asymptotic form of the exponential integrals can be approximated by the following expression
\begin{equation}
\mathrm{E}_{1}(2i\tau_2) - \mathrm{E}_{1}(2i\tau_1) \simeq
- \frac{i}{2} 
\left[
\frac{e^{-2i \tau_2}}{\tau_2}
- \frac{e^{-2i\tau_1}}{\tau_1}
\right] \, .
\label{Ei-asymptotic}
\end{equation}
Next, recall that $\tau_1$ and $\tau_2$ are close to each other. 
Then after some more algebra, one arrives at the following expression for the scalar energy-momentum tensor
\begin{equation}
\mathrm{T}^{(\phi)}_{ij} (\tau_2 , \mathbf{k} ) \simeq  
(2 \pi )^{3} \frac{2k_i k_j}{k}
\left[
\frac{2 {\tau_1}^3 + 2 (\tau_2 - \tau_1)}{2 {\tau_1}^3 - 2 {\tau_1} - 2 (\tau_2 - \tau_1) }
\right]^2 
\left[
1 - \frac{4 {\tau_1}^3 \cos(2 i \tau_1 (\tau_2 - \tau_1))}{2 {\tau_1}^3 - 2 {\tau_1} - 2 (\tau_2 - \tau_1) }
\right]
\, .
\label{DM-em-tensor-fin}
\end{equation}
The result obtained in the equation (\refeq{DM-em-tensor-fin}) contains information about the particle production of a single sub-horizon mode $\mathbf{k}$ in the time interval $(\tau_2 - \tau_1)$. The tensor  $\mathrm{T}^{(\phi)}_{ij} (\tau_2 , \mathbf{k} )$ depends on the initial time $\tau_1$, on the components of the momentum and on the time separation between the two events. 

In order to interpret the result given in (\refeq{DM-em-tensor-fin}) from the physical point of view, we note that the meaning of $\mathrm{T}^{(\phi)}_{ij} (\tau_2 , \mathbf{k} )$ is of the energy-momentum associated to the $\mathbf{k}$ particles at $\tau_2$ with respect to the vacuum defined at $\tau_1$. Therefore, $\mathrm{T}^{(\phi)}_{ij} (\tau_2 , \mathbf{k} ) $ is a source of gravitons that results from the $\mathbf{k}$-particle production. If we denote 
by $\overline{\mathrm{T}}^{(\phi)}_{ij} (\tau_2 , \mathbf{k} ) $
the source of the gravitational field generated by the $\mathbf{k}$-scalar modes in the absence of the particle generating process, then we note that the total source at $\tau_2 $ is $ \overline{\mathrm{T}^{(\phi)}}_{ij} (\tau_2 , \mathbf{k} ) + \mathrm{T}^{(\phi)}_{ij} (\tau_2 , \mathbf{k} ) $. Then the graviton mode operator can be written as follows
\begin{align}
\mathrm{h}_{ij}(\tau_2, \mathbf{k}) & = 
h_{0,ij} (\tau_2, \mathbf{k} ) +
\overline{h}_{ij} (\tau_2, \mathbf{k} ) 
+ \frac{32 \pi^2 \tau^{\frac{3}{2}}_{2}}{H^{2}_{0}} 
\frac{k_i k_j}{k^2}
\nonumber
\\
& \times \left\{
\frac{\cos(k \tau_2 )}{\sqrt{\tau_2}}
\int d \tau_2
\left[
\frac{\sin(k \tau_2)}{\tau_2}
\left[
\frac{2 {\tau_1}^3 + 2 (\tau_2 - \tau_1)}{2 {\tau_1}^3 - 2 {\tau_1} - 2 (\tau_2 - \tau_1) }
\right]^2 
\left[
1 - \frac{4 {\tau_1}^3 \cos(2 i \tau_1 (\tau_2 - \tau_1))}{2 {\tau_1}^3 - 2 {\tau_1} - 2 (\tau_2 - \tau_1) }
\right]
\right]
\right.
\nonumber
\\
& -
\frac{\sin(k \tau_2 )}{\sqrt{\tau_2}}
\int d \tau_2
\left[
\frac{\cos(k \tau_2)}{\tau_2}
\left[
\frac{2 {\tau_1}^3 + 2 (\tau_2 - \tau_1)}{2 {\tau_1}^3 - 2 {\tau_1} - 2 (\tau_2 - \tau_1) }
\right]^2 
\left[
1 - \frac{4 {\tau_1}^3 \cos(2 i \tau_1 (\tau_2 - \tau_1))}{2 {\tau_1}^3 - 2 {\tau_1} - 2 (\tau_2 - \tau_1) }
\right]
\right]
\Bigg\}
\label{hTT-DM-F-function}
\end{align}
where $h_{0,ij} (\tau, \mathbf{k} )$ is the free graviton operator that can be read off of the equation (\refeq{physical-graviton-Fourier-exp}) and $\overline{h}_{0,ij} (\tau_2, \mathbf{k} )$ is the graviton generated by the source $ \overline{\mathrm{T}^{(\phi)}}_{ij} (\tau_2 , \mathbf{k} )$. The last term from the relation (\refeq{hTT-DM-F-function}) contains the correction to the graviton due to the process of particle production in the interval $(\tau_1 , \tau_2 )$. 
In the instantaneous vacuum at $\tau_2$, the contribution from 
$ \overline{\mathrm{T}^{(\phi)}}_{ij} (\tau_2 , \mathbf{k} )$ can be calculated from the equations (\refeq{scalar-energy-momentum-final}) and (\refeq{DM-instantaneous-mode-fct-1}) and takes the following form
\begin{equation}
\overline{h}_{ij} (\tau_2, \mathbf{k} )   = 
\frac{ \tau_{2}  \sin(2k \tau_2 )} {\pi H^{2}_{0}}
\int d^3 p \, \frac{p_i (k_j - p_j)}{k\sqrt{k|\mathbf{k} - \mathbf{p}|}}
\,
\Big\{
\mathrm{Ei}[1, i(k + |\mathbf{k} - \mathbf{p}|)\tau_2 ]
-
\mathrm{Ei}[1, i(k + |\mathbf{k} - \mathbf{p}|)\tau_2 ]
\Big\} \, .
\label{no-production-graviton}
\end{equation}
The elementary integrals from the equation (\refeq{hTT-DM-F-function}) can also be calculated exactly but their expression is not very illuminating.

\section{Discussions}

In this paper, we have derived the correction to the energy-momentum tensor of a scalar field in the de Sitter space from the particle production of a single sub-horizon mode in the WKB approximation. The result is presented in the equation (\refeq{DM-energy-momentum-t2-general}) in a general form that can be applied to any pair of vacua. Its particular form obtained for two instantaneous vacua is given in the equation (\refeq{DM-em-tensor-fin}). From that, we calculated the correction to the physical graviton operator whose complete expression was given in the equations (\refeq{hTT-DM-F-function}) and (\refeq{no-production-graviton}), respectively. 

These results were obtained by considering that the three-operator expectation values of the interacting DM-graviton fields $\langle aab \rangle$ in the interaction vacuum given by the equation (\refeq{graviton-vacuum}), correspond to 
the exchange of on-shell excitations between the DM and the graviton field. However, it is not clear if tensor $\mathrm{T}^{(\phi)}_{ij} (\tau_2 , \mathbf{k} )$ from the equation (\refeq{DM-em-tensor-fin}) contributes to the graviton tadpole diagram. Previous studies show that the vacuum expectation value of the stress tensor of the linearized gravity in the Hadamard vacuum is a constant that can be absorbed into the renormalized cosmological constant \cite{Allen:1987tz}. By using the SK formalism, the quantum correction to the linearized Einstein equations with massless, minimally scalar field was obtained. Also, it was shown there that the massless scalars have no effect on the dynamical gravitons at one loop. In \cite{Park:2015kua}, it was shown that the graviton  self-energy  derived from the massless scalar in the de Sitter space induces corrections to the gravitational potentials of a static particle. 
On phenomenological grounds, these results are at apparent tension with the ones obtained in this paper at the massless limit where one expects to obtain the largest correction from the scalar field. However, it is important to note that the comparision between the results given by the WKB for massive fields and the SK for massless fields in curved space-time, respectively, is not immediate. Indeed, even in the case of the mass parameter $m=0$, the effective mass in de Sitter (used here in the WKB approximation) is still different from zero according to the equation (\refeq{scalar-effective-mass-dS}) above. Also, in order to compare the two methods for the massless field, a more careful comparision between the quantum corrected linearized Einstein equations and the Bogoliubov map, that determine the mapping between the states at two different instants, should be done. This analysis is beyond the scope of this paper. Nevertheless, a close correspondence between the results obtaine here and the ones from the literature can be drawn by noting that the corrections generated by the terms of the form $\langle aab \rangle$ are zero if the interacting vacuum is taken to be the tensor product of the non-interacting vacua of the DM and the graviton fields, respectively, which is an additional assumption.
Concerning the application of the WKB method to the quantum field theory in curved space-time, we note that one can gain more insights about it by representing the linearized excitations in other pictures as in \cite{Rajeev:2017uwk}.

The results obtained here invite to a further analysis of the interaction between the scalar field and the gravitons in de Sitter space since it is important to see in what conditions the different approaches to the perturbative quantum excitations converge. From the point of view of the possible applications, the WKB analysis performed here could be helpful to the study of the backreactions in the semi-classical treatment of gravity, as well as to the study of the particle interactions and of the DM processes in the primordial universe. From the theoretical point of view, the same steps can be taken to derive the corrections to the physical graviton two-point functions in the WKB formalism. We hope to report on these subjects elsewhere.

\section*{Acknowledgements}

I acknowledge M. C. Rodriguez for discussions.


\end{document}